\documentclass[preprintnumbers,superscriptaddress,pre,twocolumn,longbibliography]{revtex4-2}
\usepackage{stackengine}
\usepackage{amsmath}
\usepackage[sc]{mathpazo}
\usepackage{graphicx,subfigure}
\usepackage{tabularx, booktabs}
\usepackage{geometry}
\usepackage{tikz}
\newcommand{\add}[1]{\textcolor{black}{#1}}

\begin{document}
	
	\title{Emergence of power-law distributions in self-segregation reaction-diffusion processes}
	
	\author{Jean-François de Kemmeter}
	\affiliation{Department of Mathematics and naXys, Namur Institute for Complex Systems, University of Namur, Rue Graf\'e 2, B5000 Namur, Belgium}
	\affiliation{Department of Mathematics, Florida State University, 1017 Academic Way, Tallahassee, FL 32306, United States of America}
	
	\author{Adam Byrne}
	\affiliation{School of Mathematics and Statistics, University College Dublin, Belfield, Dublin 4, Ireland}
	
		\author{Amy Dunne}
	\affiliation{School of Mathematics and Statistics, University College Dublin, Belfield, Dublin 4, Ireland}
	
	\author{Timoteo Carletti}
	\affiliation{Department of Mathematics and naXys, Namur Institute for Complex Systems, University of Namur, Rue Graf\'e 2, B5000 Namur, Belgium}
	
	\author{Malbor Asllani}
	\affiliation{Department of Mathematics, Florida State University, 1017 Academic Way, Tallahassee, FL 32306, United States of America}

\begin{abstract}
Many natural or human-made systems encompassing local reactions and diffusion processes exhibit spatially distributed patterns of some relevant dynamical variable. {These interactions, through self-organization and critical phenomena, give rise to power-law distributions, where emergent patterns and structures become visible across vastly different scales. Recent observations} {reveal power-law distributions in the spatial organization of, e.g., tree clusters and forest patch sizes. Crucially, these patterns do not follow a spatially periodic order but rather a statistical one. Unlike the spatially periodic patterns elucidated by the Turing mechanism, the statistical order of these particular vegetation patterns suggests an incomplete understanding of the underlying mechanisms.} \add{H}ere, we present a novel self-segregation mechanism, driving the emergence of power-law scalings in pattern-forming systems. The model incorporates an Allee-logistic reaction term, responsible for the local growth, and a nonlinear diffusion process accounting for positive interactions and limited resources. According to a \emph{self-organized criticality} (SOC) principle, after an initial decrease, the system mass reaches {an analytically} predictable threshold, beyond which it self-segregates into distinct clusters, due to local positive interactions that promote cooperation. Numerical investigations show that the distribution of cluster sizes obeys a power-law with an exponential cutoff. 
\end{abstract}	

\date{\today}
\maketitle

%\section{Introduction}
%

\textbf{Introduction}\label{sec:I}.- Nature exhibits various forms and shapes of order, spanning from the collective flight of birds in flocks~\cite{bialek2012statistical} to the synchronized flashing of fireflies~\cite{firefly_sync}. Self-organization has long been recognized as the fundamental principle driving the emergence of such captivating patterns~\cite{ball_self-made_1999, murray_mathbio,cross_pattern_2009}. 
Exploring how these collective behaviors and patterns arise from the interactions among the system's basic units has been a vibrant research field for a long time.
Notably,  the past two decades have witnessed a growing interest in understanding the formation of vegetation patterns in semi-arid ecosystems~\cite{lefever_origin_1997, klausmeier1999regular, kefi2007spatial, Rietkerk, Bonachela, zhang2020regular, scanlon2007positive, huang_microclimate_2021,torquato2023self,ge2023hidden,hu2023disordered} where, even in harsh environmental conditions, plants manage to survive by clustering together. \add{In the following, we shall consider vegetation {patterning} as a prototype system, primarily focusing on the analysis within a broader context.} 
\add{In particular, we focus on the so-called irregular patterns \cite{klausmeier1999regular}, which lack apparent spatial order. By considering the feedback between plant biomass and resources (e.g., water), Klausmeier proposed a reaction-diffusion model capable of reproducing the emergence of spatially regular patterns, such as stripes of vegetation on hillslopes, by following a Turing-like instability  \cite{klausmeier1999regular, zhang2020regular}. This model also predicts the emergence of irregular patterns, it was suggested that they arise from the amplification of small topographic variations or quenched disorder of some other nature \cite{zimmermann1993effects,yizhaq2016effects}.}
While regular patterns, understood through Turing-like instabilities~\cite{lefever_origin_1997, Rietkerk, Bonachela, meron_vegetation_2019}, provide a foundational context with their characteristic scale length and spatial order, {our emphasis shifts towards} irregular patterns. These consist of clusters with diverse sizes, distributed seemingly at random and interspersed with bare areas~\cite{klausmeier1999regular}, {representing a more recent area of investigation}. Despite apparent disorganization at smaller scales, irregular patterns reveal an emergent global order. {For instance,} studies across various geographical regions have demonstrated that cluster size distributions in these patterns {exhibit a power-law behavior, often with an exponential cutoff~\add{\cite{scanlon2007positive,kefi2007spatial,huang_microclimate_2021,villegas2024evidence}}}, {highlighting} the {lack of complete} understanding of the {formation of} irregular patterns. \add{Differently from Klausmeier's approach, the organization of irregular patterns here is not affected by the nonuniformity of the spatial support, but rather by the positive feedback between plants and the finite size effects.}
%\del{Vegetation patterns in semi-arid regions were also found to be disordered hyperuniform, i.e., to exhibit long-range hidden order \cite{torquato2023self,ge2023hidden,hu2023disordered}. Hyperuniformity has gained increasing attention these last years across sciences. In particular, disordered hyperuniform states of matter are characterized by suppressed large-scale density fluctuations \cite{torquato2018hyperuniform}. }

\add{In t}his work\add{, we} introduce a novel self-segregation process relying on \textit{self-organized criticality} (SOC) \cite{pascual2005criticality}. The latter has proved successful in explaining emergent phenomena characterized by power-law scaling in various scenarios, such as avalanches in the sand-pile model~\cite{BTW}, forest-fire dynamics~\cite{forest_fire}, the spread of infections in epidemics~\cite{rhodes1997critical} \add{or in tropical geometry \cite{kalinin2018self}}. SOC models are distinguished by their critical state, wherein system dynamics reach a critical point as a specific dynamical variable, such as mass~\cite{BTW} or energy~\cite{Zhang_energy}, surpasses a certain threshold, instead of relying on fine-tuning of some model parameter.
By considering fundamental principles that describe individual efforts to survive against hostile (environmental) factors, as well as their dispersal in the spatial domain while considering limited resources and cooperation, we derive a reaction-diffusion equation that governs the temporal evolution of density in the spatial domain. {Notably, the deterministic reaction-diffusion process that we propose exhibits a self-organized criticality with the emergence of power-law scalings at the critical point.} To address the challenges of survival in a harsh environment, we utilize a modified logistic equation with an Allee effect~\cite{allee_1932, courchamp2008allee}. The latter models the fact that a species can only persist if its local population exceeds a specific threshold, otherwise leading to extinction. Notably, the nonlinear diffusion model developed in this study is reminiscent of earlier nonlinear random walk processes introduced by the authors {in the case of network structures}~\cite{Asllani_PRL, Carletti_PRR, Siebert_2022, DeKemmeter} {and where it has been shown that heterogeneity and social affinity lead to self-segregation, with individuals clustering in high-degree nodes, leaving the lower-degree nodes empty.} In contrast, the present study {shifts focus to continuous spatial domains; moreover we emphasize how asymmetric cooperation among neighboring individuals helps to surpass the Allee threshold, by fostering the emergence of clusters of occupied territories interspersed by vacant areas.} Analysis of cluster size distributions reveals a power-law behavior with an exponential cutoff at larger sizes.

%The paper is organized as follows. In section~\ref{sec:ILM}, we introduce the non-linear reaction-diffusion process governing the spatio-temporal evolution of the vegetation densities. In section~\ref{sec:III}, we show how this process leads to the emergence of irregular clusters separated by empty areas. We then analyze in section~\ref{sec:IV} the distribution of the cluster sizes. We summarize our findings and conclude in section~\ref{sec:V}.  

%%%%%%%%%%%%%%%%%%%%%%%%%%%%%%%%%%%%%%%%%%%%%%%%%%%%%%%%%%%%%%%%%%%%%%%%%%%%%%%%%%%%%%%%%%%%%%%%

\textbf{Individual-based model and mean-field limit}\label{sec:II}.- {We} start by considering the spatial domain $\mathcal{R}$, where the interactions between agents occur, to be a two-dimensional square support of unit length with periodic boundary conditions divided into $\Omega = L\times L$ spatial compartments or squared patches of equal area, labeled $v_{i}$ for $i=1,\cdots,\Omega$. 
%% Limited carrying capacity: plants and vacancies
{For simplicity, w}e assume each patch contains the same limited amount of generic resources, which sets the maximal number $N$ of individuals the patch can host simultaneously. The number of individuals within patch $v_i$ at time $t$ is denoted by $0\leq n_i(t)\leq N$ and thus $N-n_i(t)$ quantifies the vacancies, i.e., the additional number of individuals the patch $v_i$ might host. 
The stochastic nature of the processes at play can be modeled by using the master equation 
\begin{equation}
\frac{d \mathrm{P}(\mathbf{n},t) }{dt} =  \sum_{\mathbf{n'} \neq \mathbf{n}} 
\mathrm{T}(\mathbf{n} \vert \mathbf{n'}) \mathrm{P}(\mathbf{n'},t)
-\mathrm{T}(\mathbf{n'} \vert \mathbf{n}) \mathrm{P}(\mathbf{n},t)
\label{eq:MEQ}
\end{equation}
which provides a detailed probabilistic description of the dynamics starting from the microscopic setting. Here $\mathbf{n}(t) = \left(n_1\left(t\right),n_2\left(t\right),\cdots,n_{\Omega}\left(t\right)\right)$ is the state vector and $\mathrm{P}(\mathbf{n},t)$ is the probability that the system will be in such a state {at time $t$}. Furthermore $\mathrm{T}(\mathbf{n'} \vert \mathbf{n})$ denotes the transition probability, per time unit, from state $\mathbf{n}$ to state $\mathbf{n}'$ and the summation in Eq.~\eqref{eq:MEQ} extends over all the states different from $\mathbf{n}$. 
%%
%% Dynamics
We will assume that individuals interact with each other both within each patch and between adjacent ones. The dynamics at a purely local level will capture the natural death process for which an agent will be removed from the $i$-th patch, $X_{i} + E_{i} \xrightarrow{r_1} 2 E_{i}$, where $X_{i}$, $E_{i}$, and $r_1$ denote a single individual, a single vacancy and death rate, respectively. On the other side, the birth process of an agent in any patch $i$, is constrained by a strong Allee effect \cite{courchamp2008allee}, i.e., $2 X_{i} + E_{i} \xrightarrow{r_2} 3X_i$, with a birth rate $r_2>r_1$ to allow survivability. 
The finite carrying capacity encapsulates not simply limited resources but all other possible factors with a negative impact on the growth and survivability of the species, such as the presence of predators, intra- or inter-species competition, lack of potential mating partners and so on, broadly known as the Allee effect~\cite{allee_1932,stephens1999allee,courchamp1999inverse,dennis2002allee,taylor2005allee,courchamp2008allee,
kindvall1998individual,kanarek2015overcoming,Asllani_Allee}.
In conclusion, the dynamics at the level of patch $i$ will be described by the following transition rates
\begin{subequations}
	\begin{align}
	&T(n_i-1 \vert n_i) = \frac{r_1}{\Omega} \dfrac{n_i}{N}\Big(1-\dfrac{n_i}{N}\Big),\\
	&T(n_{i}+1 \vert n_{i}) = \frac{r_2}{\Omega}\dfrac{n_i}{N}\dfrac{n_i-1}{N}\Big(1-\dfrac{n_i}{N}\Big),
	\end{align}
	\label{eq:BirthDeath}
\end{subequations}
for the death and birth dynamics, respectively. 
 
On the other side, the individuals are allowed to interact with each other at the inter-patch level, i.e., $X_{i} + aX_{j} + E_{j} \xrightarrow{\delta} E_{i} + 2X_{j}$ with $a>0$, where the previous reaction, occurring with a rate $\delta$, models the process by which a plant sends its seed to a neighboring patch before dying. Such an interaction is the {key} point of this paper and  describes the asymmetric {mutualistic} interaction between individuals of different patches while taking into account the finite carrying capacity of each site \citep{fanelli_mckane, Asllani_PRL, Carletti_PRR}. 
Provided $v_i$ and $v_j$ are neighbor sites, the transition from $v_i$ to $v_j$ reads
\begin{equation}
T(n_{i}-1,n_{j}+1 \vert n_i, n_j) 
= \frac{\delta}{k \Omega} \frac{n_i}{N} {\Big(\frac{n_j}{N}\Big)}^a {\Big(1-\frac{n_j}{N}\Big)},
\label{eq:ratesDiff}
\end{equation}	
with $k$ the number of neighbors per site, i.e., $k=4$ in the present setting.
The dispersion of the vegetation in the spatial domain will thus act as a trade-off between the positive interactions between  individuals and the finite carrying capacity. {Studies have shown that positive spatial feedbacks, such as improved water retention under tree canopies, support tree growth and survival, yet the ecosystem's limited resources, dictated by rainfall and nutrients, keep these dynamics under control, avoiding overgrowth or desertification \cite{wang_biosphereatmosphere_2000,scholes_tree-grass_1997,nathan_mechanisms_2002,Molofsky2001,Schlesinger1996}.} In the following, we will assume $a>1$. This requirement implies an asymmetry in the interaction between individuals of adjacent sites, i.e., they will perceive a {higher} number of individuals than those available on the hosting site. Inspired by the ecological literature~\cite{Weiner}, we will refer to it as \emph{size-asymmetric interaction}. 

Starting from the master equation \eqref{eq:MEQ} we will look for a mean-field formalism, see SM \footnote{See Supplemental Material, which contains Refs. \cite{mckane2004stochastic,pruessner2012self} for more details about the individual-based modeling and averaging method, the fixed points of the system and their stability, the slow-fast limit and intermittency of the system}. 
%for which a detailed derivation can be found in the Appendix~\ref{app:ME}. 
Let us here recall that the standard approach is to consider the time evolution of the {density of agents} $\langle n_i\rangle/N$ within the site $v_i$ in the limit $N \rightarrow +\infty$ and then take the continuum limit in which the {number of} mesh {points} goes to {infinity, i.e., $\rho=\lim_{N \to +\infty, L \to +\infty} {\langle n_i\rangle}/{N}
$}. This procedure leads to the following partial differential equation for the time evolution of species density $\rho\equiv \rho(\mathbf{x},t)$ at point $\mathbf{x}=(x,y)$ and time $t$: 
\begin{equation}
\frac{\partial \rho}{\partial t}=
r\, f(\rho)
+ D \, \Big[
g{(\rho)}\Delta \rho - \rho \Delta g{(\rho)}
\Big]\ .
\label{eq:PDE}
\end{equation}
Here $D>0$ represents the diffusion coefficient, $\Delta{=\partial_x^2+\partial_y^2}$ the Laplace operator, $f(\rho)=\rho(1-\rho)(\rho-A)$ the Allee reaction term with $r>0$ the growth rate and $0<A<1$ the Allee coefficient \footnote{In the SM \cite{Note1} we show how to determine the expressions for the coefficients $D, A$ and $r$ as a function of the rates above introduced. More precisely, the diffusion coefficient is obtained as the limit $D = \lim_{\Omega\rightarrow \infty} \delta/\Omega$. Similarly, the growth rate is obtained as the limit $r = \lim_{\Omega\rightarrow \infty} r_2/\Omega$. Finally $A=r_1/r_2$}. The function $g(\rho)=\rho^a(1-\rho)$ captures in a compact form the nonlinear interacting terms between individuals of neighbor sites. Let us observe that if $r=0$ the total mass is conserved (see SM~\cite{Note1}). %\newline

\begin{figure*}[htb!]
	\centering
	\includegraphics[width=\textwidth]{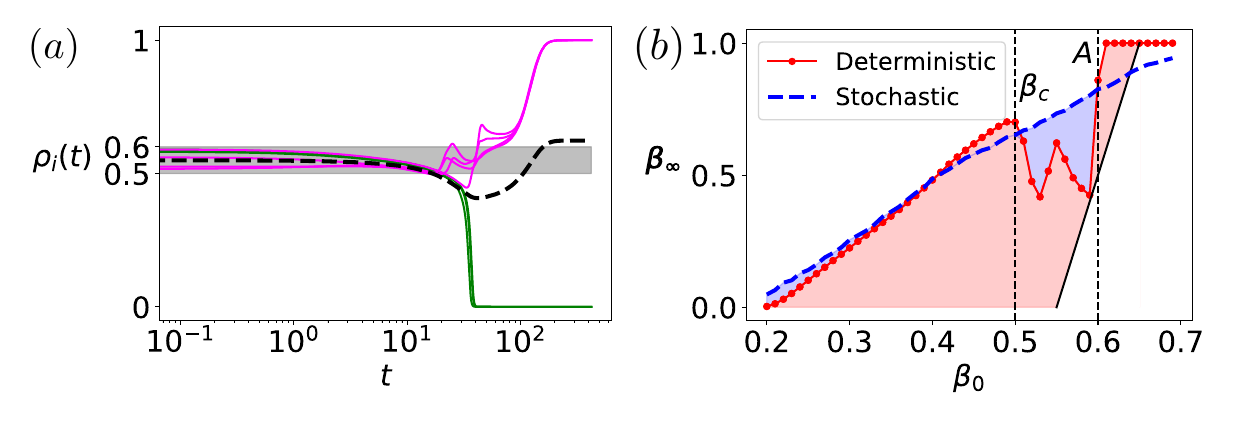}
    \caption{$\textbf{(a)}$ Site densities evolve in a square lattice, starting from uniform densities in $[0.5,0.6]$. As densities cross $\beta_c=0.5$, self-segregation occurs, with some reaching carrying capacity (pink, $\rho=1$) and others (green) converging to $\rho=0$. Average node density evolution is shown by the dashed black line.
    $\textbf{(b)}$ Final average density, $\beta_{\infty}$, versus initial density, $\beta_0$, {is depicted} with red points{, and the fraction of nodes above the Allee constant is represented by a black line for comparison.} Shaded red region indicates survivability gain from self-organized criticality (SOC). Blue dashed line and shaded region show stochastic system superiority over deterministic (averaged over $10$ Gillespie algorithm realizations). Parameters: $A=0.6$, $r=1/6$, $D=10$, $a=2$. Spatial mesh: $500 \times 500$ points. Initial conditions: $[\beta_0-0.05, \beta_0+0.05]$, rescaled for initial average density $\beta_0$.}
\label{fig:SOC}
\end{figure*}

%%%%%%%%%%%%%%%%%%%%%%%%%%%%%%%%%%%%%%%%%%%%%%%%%%%%%%%%%%%%%%%%%%%%%%%%%%%%%%%%%%%%%%%%%%%%%%%%

\textbf{Self-segregation process as a self-organized criticality mechanism}\label{sec:III}.- As a preview of our findings, we will establish that irregular patterns arise when the total mass of the system reaches a critical value, a characterizing feature of SOC processes \cite{pascual2005criticality}. Let us first observe that a uniformly distributed density $\rho(\mathbf{x},t)=\beta$ with $\beta={0,A,1}$ represents a stationary solution of Eq. \eqref{eq:PDE}. The stability of these states can be determined by analyzing the linear evolution of the perturbation $\delta \rho(\mathbf{x},t)$, governed by the equation
\begin{equation}
\frac{\partial \delta \rho}{\partial t} = f'(\beta)\delta \rho + D\left[g(\beta)-\beta g'(\beta)\right] \Delta \delta \rho\, .
\label{eqA:Stab}
\end{equation}
It can be readily verified that $f'(\beta)<0$ when $\beta={0,1}$ and $f'(\beta)>0$ when $\beta=A$, indicating the bistable nature of the Allee model. By seeking solutions of the form $\delta \rho \sim \sum_{\mathbf{k}} e^{\lambda_{\mathbf{k}}t}e^{i\mathbf{k}\cdot \mathbf{x}}$, we obtain the dispersion relation
\begin{equation}
\lambda_{\mathbf{k}}=f'(\beta)-D \left[g(\beta)-\beta g'(\beta)\right]|\mathbf{k}|^2\, ,
\label{eq:disp_rel}
\end{equation}
where $|\mathbf{k}|^2=k_1^2+k_2^2$ is the {square of the} module of the vector $\mathbf{k}$. For the fixed point $\beta=A$, there will always exist a finite interval (e.g., near the origin {where $|\mathbf{k}|^2$ is small enough}) for which $\lambda_{\mathbf{k}} > 0$, proving its unstable behavior. Conversely, the other homogeneous fixed points $\beta={0,1}$ are stable as long as the effective diffusion coefficient $D_{\mathit{eff}}=D\left[g(\beta)-\beta g'(\beta)\right]$ is non-negative, a condition that holds true but that does not contribute to pattern formation because they will {represent} global extinction or a fully occupied domain. {Furthermore}, the dynamics stemming from the unstable state $\beta=A$ could not guarantee the emergence of any {nontrivial} spatial pattern organized into separate clusters {as the system might converge to the fully occupied or empty state}. Eq.~\eqref{eq:PDE} displays other stationary solutions, whose existence and stability are addressed in the following, by adopting an approach based on slow-fast dynamics. Specifically, we consider the limit $r/D\rightarrow 0$, where the fast dynamics is solely governed by the nonlinear diffusion process. {Let us observe that this separation of timescales is in line with SOC \cite{pascual2005criticality}.} In general, diffusion processes tend to homogenize the spatial distribution of mass. However, as previously {mentioned}, under certain conditions, the effective diffusion coefficient can become negative ($D_{\mathit{eff}}<0$). Negative diffusion exhibits the opposite effect of homogenization, leading to the accumulation and localization of mass within the spatial domain~\cite{karpov_negative_1995,argyrakis_negative_2009}.
Motivated by this insight, we first note that, in contrast to the full reaction-diffusion equation, the nonlinear diffusion operator vanishes for every uniform state $\rho(\mathbf{x})=\beta>0$. At this stage, we can ascertain the critical value $\beta_c$ of the average node density $\beta$ {below} which the equilibrium $\rho(\mathbf{x})=\beta$ undergoes instability due to diffusion. It can be easily shown (see SM~\cite{Note1}), that this critical value is given by
\begin{equation}
\beta_c=\frac{a-1}{a}\, .
\label{eq:beta_c}
\end{equation}
This formula justifies the choice of $a > 1$ for heterogeneous patterns to develop, i.e., the asymmetry in the interactions along with the cooperation between individuals of adjacent sites allows for the self-segregation to occur{; indeed if $a<1$, then $\beta_c<0<\beta$, returning $\rho(\mathbf{x})=\beta$ to be a stable homogeneous solution}. Any uniform state $\rho(\mathbf{x})=\beta < \beta_c$ becomes unstable, while it remains stable otherwise. Upon instability, due to mass conservation, a redistribution of mass is expected to occur. The latter takes place in the form of clusters, hereby referred to as connected subregions of homogeneous mass, separated by empty patches. The size of the cluster is then defined as the {the contiguous area covered, numerically calculated as the} number of {connected} patches it contains. For a cluster $\mathcal{C}_s$ to be stable, the local density must satisfy $\beta_{\mathit{loc}}=\frac{1}{\vert \mathcal{C}_s \vert }\int_{\mathcal{C}_s} \rho (\mathbf{x}) \text{d}\mathbf{x} >\beta_c$, where $\vert \mathcal{C}_s \vert$ is the area of the cluster $\mathcal{C}_s$. Once the (fast) diffusion creates a precursor of what will become a stable uniform cluster, the (slow) reaction comes into play by maximizing the cluster density to unity if $\beta_{loc}>A$ or reducing it to zero otherwise~\footnote{In addition, considering the physics of the problem, each cluster adheres to no flux boundary conditions $\nabla \rho(\mathbf{x})=0$.}. This approach uncovers the presence of heterogeneous (stable) stationary solutions. A more comprehensive and rigorous proof is provided in the SM~\cite{Note1}.
This initial finding unveils a fundamental insight: the stability of a specific state {crucially} relies on the total mass of the state itself. This property aligns perfectly with the concept of self-organized criticality (SOC), which pertains to the inherent self-organization of a system when it reaches a critical threshold of a globally defining observable, such as mass or energy~\cite{BTW, Zhang_energy}.

\begin{figure*}[htb!]
	\centering
		\includegraphics[width=\textwidth]{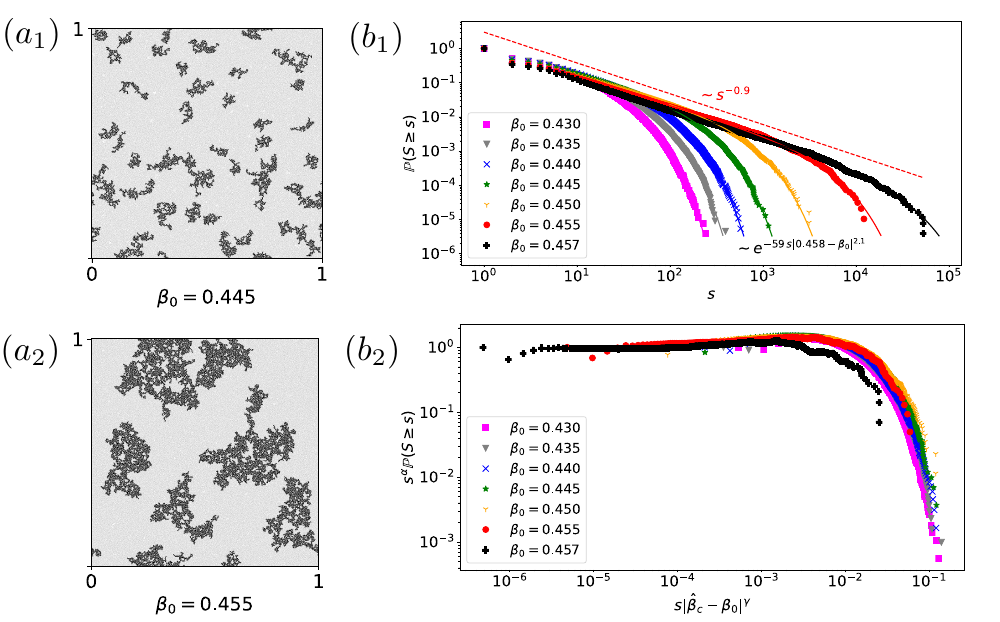}
		\caption{$\textbf{(a)}$ Vegetation patterns for two average initial site density values: (upper) $\beta_0=0.445$ and (lower) $\beta_0=0.455$.
$\textbf{(b)}$ {(Upper)} Cumulative distribution of cluster sizes, each curve for a distinct $\beta_0$ (results from $10$ configurations). Data align with a power-law distribution (red dashed curve) with an exponential cutoff. ({Lower}) Data collapse onto a universal curve by plotting $s^{\alpha} \mathbb{P}(S \geq s)$ against $s{\vert \hat{\beta}_c - \beta_0 \vert}^{\gamma}$. {Deviations in the black curve from exponential fitting stem from finite size effects, manifesting at large $s$.} Initial node densities were in $[\beta_0-0.05, \beta_0+0.05]$, rescaled for initial average density $\beta_0$. Simulations on a $500\times 500$ square lattice, parameters: $a=2$, $A=1/6$, $r=0.1$, $D=10$.}
		\label{fig:PowerLaw}
\end{figure*}

In Fig.~\ref{fig:SOC}, panel $(a)$, we consider a slow-fast dynamics setting, i.e., $r/D \ll 1$, where initially, the total mass is randomly distributed throughout the spatial domain, yet always below the Allee threshold and above the critical value $\beta_c$. The strong diffusion tends to initially homogenize the mass, which decreases since it remains below the survivability threshold. As expected in SOC dynamics, when the global observable (density in our case) reaches the critical value $\beta_c$, a change of behavior occurs, leading to an overall increase in mass across most lattice sites. The remarkable {and conterintuitive} aspect is that the species manage to survive in the {stationary} state{, i.e., $\lim_{t\rightarrow +\infty} \rho_i(t) = 1$ for some nodes $i$}, although the {initial} density at each site is below the Allee parameter, as {confirmed} by the {early} trend. As the average density decreases further, a new phenomenon emerges: self-segregation. Driven by the negative value of $D_{\mathit{eff}}$, the mass rapidly accumulates and localizes in different subregions of the domain $\mathcal{R}$. If the densities of the new clusters surpass both the critical values of self-segregation and Allee ($\beta_{\mathit{loc}} > \beta_c, A$), the species will survive in those particular clusters and eventually reach a full carrying capacity $\beta_{\mathit{loc}}=1$, as illustrated in Fig.~\ref{fig:SOC}, panel $(a)$. {I}n the SM \cite{Note1} we {also} give evidence of intermittency, a characterizing feature of SOC models. %\JFadd{As shown in Appendix \ref{app:Intermittency} and similarly to the sandpile model \cite{pruessner2012self}, this self-segregation mechanism is characterized by intermittent events, i.e., the formation of clusters takes place through avalanches interspersed by long time periods in which the number of empty nodes remains constant. }
The benefit of self-segregation for individual survivability is systematically investigated in Fig.~\ref{fig:SOC}, panel $(b)$, where various initial density values $\beta_0=\int_{\mathcal{R}}\, \rho(\mathbf{x},0)\mathrm{d}\mathbf{x}$ are considered. In all cases, the species survive beyond intuitive expectations. Particularly, in the interval $\beta\in[\beta_c, A]$, the diffusion has a homogenizing effect by reducing the initial perturbation, {thereby} slowing down the fast dynamics of the diffusion component. Consequently, the final equilibrium density is lower than the initial density. However, this outcome is an artifact of the deterministic mean-field approach utilized here. In a real scenario, the presence of external or demographic noise acts as a permanent perturbation (forcing) term, preventing a substantial decrease in the final density compared to the critical value $\beta_c$. Stochastic simulations, performed using the Gillespie algorithm, are depicted by the blue dashed line (and corresponding shaded blue region) in Fig. \ref{fig:SOC} (b), thereby substantiating our claim.

%%%%%%%%%%%%%%%%%%%%%%%%%%%%%%%%%%%%%%%%%%%%%%%%%%%%%%%%%%%%%%%%%%%%%%%%%%%%%%%%%%%%%%%%%%%%%%%%

\textbf{Power-law distribution in  self-segregation patterns}\label{sec:IV}.- SOC processes are renowned for the presence of power-law distributions of some relevant variables. This is the case for instance of the sandpile model where the size of generated avalanches has a scale-free distribution \cite{BTW}. Based on the intuition that in the present model the relevant variable will be the cluster size, we have conducted a significant number of independent simulations of Eq.~\eqref{eq:PDE} with various initial values for the density $\beta_0$, closer and closer to the critical value of the system for which patterns are expected to emerge. In the slow-fast setting, this critical value is anticipated to be close to $\beta_c$. Fig.~\ref{fig:PowerLaw}, panels $(a_1)$ and $(a_2)$, show patterns with clusters of varying sizes for two different {values of the initial density $\beta_0$}. 
In Fig.~\ref{fig:PowerLaw} {$(b_1)$} we show the cumulative distribution $\mathbb{P}(S \geq s)$ of the size $S$ of the stationary clusters resulting from Eq.~\eqref{eq:PDE}. It can readily be observed that they fit very well to a power-law function with {almost} the same critical exponent $\alpha$ and are characterized by different values of exponential cutoffs {that depend on} the initial density $\beta_0$. In summary
\begin{equation}
\mathbb{P}(S \geq s) {=\kappa} s^{-\alpha}e^{s\,\xi(\beta_0)}\,,
\end{equation}
where {$\kappa$ is a normalization constant and} the function $\xi(\beta_0)$ vanishes when $\beta_0$ equals $\hat{\beta}_c$, the value for which a perfect power-law relation is observed. Inspired by similar scenarios as {in} the Ising model~\cite{sethna_statistical_2021} or the percolation processes~\cite{ding2014numerical}, we set $\xi(\beta_0) {=-C} {\vert \hat{\beta}_c - \beta_0 \vert}^{\gamma}${, with $C>0$ being an appropriately chosen scaling constant, leading} to a second exponent $\gamma>0$ which describes the transition to a genuine power-law and thus the independence of the exponential cutoff from the size of the system. In Fig.~\ref{fig:PowerLaw} {$(b_1)$} we have shown with a red dashed line and colored solid lines, respectively, for the power-law and the exponential cutoff, the best fit to the empirical critical exponents $\alpha \approx 0.9$ and $\gamma \approx 2.1$. Let us observe that both fits agree well with the numerical data except for small {and large} values of $s$ due to the finite resolution of the numerical simulations. Since the latter two parameters are independent of the values of the initial system mass suggests that our model belongs to {a} universality class, typical of systems where power-law distributions emerge~\cite{sethna_statistical_2021}. A compact way to illustrate this is by plotting $s^\alpha \mathbb{P}(S \geq s)$ as a function of $s\,{\vert \hat{\beta}_c - \beta_0 \vert}^{\gamma}$; the different curves now collapse onto a single one, known in the literature as the universal curve, {shown in} panel {$(b_2)$}. In conclusion, we assert that while the universal power-law distribution of patch sizes is solely driven by the self-segregation process (see SM~\cite{Note1}), the reaction component is vital for accurately describing the resilience of individuals in harsh conditions.%\newline

%%%%%%%%%%%%%%%%%%%%%%%%%%%%%%%%%%%%%%%%%%%%%%%%%%%%%%%%%%%%%%%%%%%%%%%%%%%%%%%%%%%%%%%%%%%%%%%%

\textbf{Conlusions}\label{sec:V}.- In this study, we presented a novel dynamical model that addresses the emergence of spatially-extended patterns, characterized by a power-law distribution of spatial cluster sizes. By considering positive interactions between individuals and accounting for limited resources, we develop{ed} a self-consistent mathematical formalism. The model encompasses a single-species evolution equation with a local reaction term based on the Allee-logistic function. To capture the spatial dynamics, we introduced a nonlinear diffusion term that models the phenomenon of self-segregation. The latter process assumes a critical role in initiating pattern formation and establishing a mechanism of self-organized criticality. Within this framework, we observe an initial decrease in mass, driven by insufficient resource availability, until a threshold, that can  be analytically predicted, is reached. Beyond this threshold, we observe the spatial organization of mass into distinct clusters characterized by higher densities, thus fostering cooperative behaviors among individuals. Consequently, clusters with densities surpassing the Allee threshold shape the final pattern. Numerical investigations confirm that the distribution of cluster sizes follows a power-law function with an exponential cutoff.% \del{This model establishes a foundation for understanding the self-organizing criticality mechanism underlying power-law distributions in spatial patterns, paving the way for new directions in pattern formation research.}

\textbf{Acknowledgments} This work was supported by the Fonds de la Recherche Scientifique-FNRS under the grant FC 38477 to J.-F. dK. Part of the results were obtained using the computational resources provided by the “Consortium des Equipements de Calcul Intensif” (CECI), funded by the Fonds de la Recherche Scientifique de Belgique (F.R.S.-FNRS) under Grant No. 2.5020.11 and by the Walloon Region.
\nocite{mckane2004stochastic}
\nocite{pruessner2012self}

\bibliographystyle{apsrev4-2}%{abbrv}%{unsrtnat}%{plainnat}%{abbrvnat}%{apsrev4-2}
\bibliography{biblioV2}% Produces the bibliography via BibTeX.

\clearpage
\onecolumngrid
\appendix

\section*{Supplemental material}

\section{Mathematical details of the individual-based modeling}\label{app:ME}

\subsection{Transition rates}
This section aims to derive the transition probabilities given by Eq.~\eqref{eq:BirthDeath}-\eqref{eq:ratesDiff} in the main text from the corresponding reactions:
\begin{subequations}
	\begin{align}
	X_{i} + E_{i} \xrightarrow{r_1} 2 E_{i} 
	\quad &\Longrightarrow \quad T(n_{i}-1 \vert n_i) = \frac{r_1}{\Omega} \frac{n_i}{N}\frac{N-n_i}{N}\label{eqA:rateDeath}, \\
	2 X_{i} + E_{i} \xrightarrow{r_2} 3X_i  
	\quad &\Longrightarrow \quad T(n_{i}+1 \vert n_{i}) = \frac{r_2}{\Omega} \frac{n_i}{N}\frac{n_i-1}{N}\frac{N-n_i}{N} \label{eqA:rateBirth},\\
	X_{i} + aX_{j} + E_{j} \xrightarrow{\delta} E_{i} + 2X_{j} 
	\quad &\Longrightarrow \quad T(n_{i}-1,n_{j}+1 \vert n_i, n_j) = \frac{\delta}{\Omega k} \frac{n_i}{N} g\Big(\frac{N-n_j}{N}\Big),\\
	X_{j} + aX_{i} + E_{i} \xrightarrow{\delta} E_{j} + 2X_{i}  
	\quad &\Longrightarrow \quad T(n_{i}+1,n_{j}-1 \vert n_i, n_j) = \frac{\delta}{\Omega k} \frac{n_j}{N} g\Big(\frac{N-n_i}{N}\Big)\, ,
	\end{align}
	\label{eqA:rates}
\end{subequations}
where $g(\rho)=\rho^a(1-\rho)$. 

The derivation is similar to the one discussed in~\cite{mckane2004stochastic}. As a preliminary step, let us remind how to compute the probability $\mathcal{P}(X=k,E=\ell)$ to pick without reinsertion $k\leq n$ letters $X$ and $\ell\leq N-n$ letters $E$ in an urn that contains $n$ letters $X$ and $N-n$ letters $E$. As a first step, let us determine the probability of picking $k$ consecutive letters $X$ followed by $\ell$ consecutive letters $E$. This probability reads:
\begin{equation}
\Big(\frac{n}{N}\frac{n-1}{N-1}\cdots\frac{n-k+1}{N-k+1}\Big)
\Big(\frac{N-n}{N-k}\frac{N-n-1}{N-k-1}\cdots\frac{N-n-\ell+1}{N-k-\ell+1}\Big)
= \frac{n!}{(n-k)!} \frac{(N-k-\ell)!}{N!} \frac{(N-n)!}{(N-n-\ell)!}\,.
\end{equation}
The probability $\mathcal{P}(X=k,E=\ell)$ is then obtained by multiplying the above expression by the number of distinct configurations obtained upon permutations of $X$ and $E$, which is given by the binomial coefficient $\binom{k+\ell}{\ell}$. Overall, we obtain:
\begin{equation}
\mathcal{P}(X=k,E=\ell) = \frac{\binom{N-n}{\ell}\binom{n}{k}}{\binom{N}{k+\ell}}\,.
\end{equation}
It follows that for~\eqref{eqA:rateBirth}, for instance, the probability to pick two agents $X$ and one vacancy $E$ within node $v_i$ is given by $3 \dfrac{n_i}{N}\dfrac{n_i-1}{N-1}\dfrac{N-n_i}{N-2}$. By denoting by $p_2$ (resp. $p_1$) the probability that the reaction will be of type~\eqref{eqA:rateBirth} (resp.~\eqref{eqA:rateDeath}) and using the fact that node $v_i$ is selected with probability $1/\Omega$, the transition rate corresponding to reaction ~\eqref{eqA:rateBirth} is given by:
\begin{equation}
\begin{split}
T(n_i+1 \vert n_{i}) &\, \sim \, 3 r_2 \frac{p_2}{\Omega}\frac{n_i}{N}\frac{n_i-1}{N-1}\frac{N-n_i}{N-2} \, \sim \, 3 r_2 \frac{p_2}{\Omega} \frac{N}{N-1} \frac{N}{N-2} \frac{n_i}{N}\frac{n_i-1}{N}\frac{N-n_i}{N}\\
&\equiv \frac{r_2'}{\Omega} \frac{n_i}{N}\frac{n_i-1}{N}\frac{N-n_i}{N},
\end{split}
\end{equation}
with $r_2'$ proportional to $r_2$. Without affecting the results in the paper, one can omit the $'$ notation in the above reaction rate (which amounts to relabeling $r_2'$ into $r_2$).
Similarly, one obtains:
\begin{equation}
\begin{split}
T(n_i-1 \vert n_{i}) &\, \sim \, 2 r_1 \frac{p_1}{\Omega} \frac{n_i}{N}\frac{N-n_i}{N-1} \, \sim \, 2 r_1 \frac{p_1}{\Omega} \frac{N}{N-1} \frac{n_i}{N}\frac{N-n_i}{N}\\
&\equiv \frac{r_1'}{\Omega} \frac{n_i}{N}\frac{N-n_i}{N},
\end{split}
\end{equation}
with $r_1'$ proportional to $r_1$. As before, one can omit the $'$ notation. With probability $1-p_1-p_2$, the reaction will correspond to the displacement of an agent between neighboring sites. Since the probability to select node $v_i$ and one of its neighbors, $v_j$ is given by $\frac{1}{\Omega k}$ with $k=4$ (each node has four nearest neighbors), we obtain:
\begin{equation}
\begin{split}
&T(n_{i}-1,n_{j}+1 \vert n_i, n_j) \, \sim \, \delta (1-p_1-p_2) \frac{1}{\Omega k} \frac{n_i}{N} g\Big(1-\frac{n_j}{N}\Big) \equiv \frac{\delta'}{\Omega k} \frac{n_i}{N} g\Big(1-\frac{n_j}{N}\Big),\\
&T(n_{i}+1,n_{j}-1 \vert n_i, n_j) \, \sim \, \delta (1-p_1-p_2) \frac{1}{\Omega k} \frac{n_j}{N} g\Big(1-\frac{n_i}{N}\Big) \equiv \frac{\delta'}{\Omega k} \frac{n_j}{N} g\Big(1-\frac{n_i}{N}\Big).
\end{split}
\label{eqA:ratesDiff}
\end{equation}
Again, one can relabel $\delta'$ into $\delta$.

One can thus rewrite the master equation as follows:
\begin{equation}
\begin{split}
\frac{d \mathrm{P}(\mathbf{n},t) }{dt}=
&\sum_{i}
\Big[
T(n_i \vert n_i+1)\mathrm{P}(n_i+1)
+  T(n_i \vert n_i-1)\mathrm{P}(n_i-1)
-  T(n_i+1 \vert n_{i})\mathrm{P}(n_i)
-  T(n_i-1 \vert n_{i})\mathrm{P}(n_i)
\Big]\\
+ &\sum_{i} \sum_{j \in \mathcal{N}(i)}
\Big[
T(n_i,n_j \vert n_i+1,n_j-1) P(n_i+1,n_j-1)
+ T(n_i,n_j \vert n_i-1,n_j+1) P(n_i-1,n_j+1)
\Big]\\
- &\sum_{i} \sum_{j \in \mathcal{N}(i)}
\Big[
T(n_i-1,n_j+1 \vert n_i, n_j) P(n_i, n_j)
+ T(n_i+1,n_j-1 \vert n_i, n_j) P(n_i, n_j)
\Big],
\end{split}
\label{eq:MEQ2}
\end{equation}
where $\mathcal{N}(i)$ denotes the set of (nearest) neighbors of node $i$. For the sake of clarity, we only highlighted the entries corresponding to the site(s) involved in the reaction. For instance, $\mathrm{P}(n_i-1)$ is the probability that the state of the system at time $t$ is given by $\mathbf{n'}= (n_1,n_2,\cdots,n_{i-1},n_i-1,n_{i+1},\cdots,n_\Omega)$.

\subsection{Details about the averaging method and mass conservation}
Let us now denote by $\langle n_i \rangle$ the average number of agents within node $v_i$, where the average is performed over all the stochastic realizations of the system.
Starting from the master equation~\eqref{eq:MEQ2}, the time evolution of $\langle n_i \rangle$ is given by:
\begin{equation}
\begin{split}
\frac{d \langle n_i \rangle}{d\tau } = 
\Bigr \langle T(n_i+1\vert n_i) \Bigr \rangle
- \Bigr \langle T(n_i-1\vert n_i) \Bigr \rangle
+ \sum_{j \in \mathcal{N}(i)} \Bigr \langle T(n_i+1, n_j-1\vert n_i, n_j) \Bigr \rangle
- \sum_{j \in \mathcal{N}(i)} \Bigr \langle T(n_j+1, n_i-1\vert n_i, n_j) \Bigr \rangle\,.
\end{split}
\end{equation}
Let us then substitute the transition probabilities by their expressions given in Eq.~\eqref{eqA:rates} and let us take the limit $N\rightarrow + \infty$. Upon rescaling of the time $t=\frac{\tau}{N}$, one finds:
\begin{equation}
\begin{split}
\frac{d \langle \frac{n_i}{N} \rangle}{dt} = 
- \frac{r_1}{\Omega} \Bigr \langle \frac{n_i}{N} \Bigr \rangle \Bigr \langle 1-\frac{n_i}{N} \Bigr \rangle 
+ \frac{r_2}{\Omega} {\Bigr \langle \frac{n_i}{N} \Bigr \rangle}^2 \Bigr  \langle 1-\frac{n_i}{N} \Bigr \rangle 
+ \frac{\delta}{\Omega} \sum_{j \in \mathcal{N}(i)} \frac{1}{k}\Bigr \langle \frac{n_j}{N} \Bigr \rangle g\Big( \Bigr \langle \frac{n_i}{N} \Bigr \rangle \Big)
- \frac{\delta}{\Omega} \sum_{j \in \mathcal{N}(i)} \frac{1}{k}\Bigr \langle \frac{n_i}{N} \Bigr \rangle  g\Big(\Bigr \langle \frac{n_j}{N} \Bigr \rangle \Big)\, .
\end{split}
\end{equation}  
Recalling the definition $\rho_i = \lim_{N\rightarrow +\infty} \Bigr \langle \frac{n_i}{N} \Bigr \rangle$, one obtains:
\begin{equation}
\begin{split}
\dot{\rho_i}=  \frac{r_2}{\Omega} \rho_i (1-\rho_i)(\rho_i - r_1/r_2)
+ \frac{\delta}{\Omega k} \sum_{j \in \mathcal{N}(i)} [\rho_j g(\rho_i)-\rho_ig(\rho_j)], ~i=1,\cdots,\Omega\,.
\label{eq:ODEs}
\end{split}
\end{equation}
The reaction part is a cubic polynomial in $\rho_i$, modeling the Allee effect. One immediately sees that any state in which sites either are fully occupied, i.e., $\rho_i^*=1$, or fully empty, i.e., $\rho_i^*=0$ will be a fixed point of the system. There are in total $2^\Omega$ of such fixed points, each of them being (locally) stable as shown in Appendix~\ref{app:stabilityMatrix}. %\JF{[So far, the computation is restricted to the case $b\geq 1, a>1$]}. 
Taking the continuum limit, i.e., $\Omega \rightarrow + \infty$ while keeping the size of the domain fixed, leads to the following partial differential equation governing the spatio-temporal evolution of the vegetation density $\rho$ 
\begin{equation}
\frac{\partial \rho}{\partial t}=
r\, \rho(1-\rho)(\rho-A) 
+ D \, \Big[
g(\rho)\Delta \rho - \rho \Delta g(\rho)	
\Big]\, ,
\label{eqA:PDE}
\end{equation}
where we have defined the positive and bounded quantities $r=\lim_{\Omega\rightarrow \infty}r_2/\Omega$ and $D=\lim_{\Omega\rightarrow \infty}\delta/\Omega$ and we assume $0<A:=r_1/r_2<1$.
%Periodic boundary conditions translate into $\rho(x=0,t)=\rho(x=1,t)$ and $\rho(y=0,t) = \rho(y=1,t)$
In the above expression, $\rho \equiv \rho(\mathbf{x},t)$ is defined on the square domain $\mathcal{R}=[0,1] \times [0,1]$ with periodic boundary conditions, $\mathbf{x}\equiv (x,y)$ and $\Delta \equiv \frac{\partial }{\partial x^2} + \frac{\partial }{\partial y^2}$.
The nonlinear diffusion (second term of the r.h.s.) preserves the total mass $M=\int_\mathcal{R} \rho(\mathbf{x},t) d\mathbf{x}$. Indeed, by assuming $r=0$, one has:
\begin{equation}
\frac{d M}{dt} = D \int_\mathcal{R} (g(\rho)\Delta \rho - \rho \Delta g(\rho)) d\mathbf{x}=0,
\end{equation}
as follows upon integrating by parts and using the assumption of periodic boundary conditions.

\section{Fixed points of the system and their stability}\label{app:fixedpoints}
In this appendix, we assume $g(x)=x^a(1-x)$ with $a>1$ and investigate the stability of the fixed points of the ODE system given by~\eqref{eq:ODEs}, namely (upon relabelling),
\begin{equation}
\dot{\rho_i}= r \rho_i (1-\rho_i)(\rho_i - A) + \frac{D}{4} \sum_{j \in \mathcal{N}(i)} [\rho_j g(\rho_i)-\rho_ig(\rho_j)], ~i=1,\cdots,\Omega \, .
\end{equation}
Let us first consider the homogeneous fixed points of the system. We will denote by $\rho_i^*$ the stationary density within site $v_i$. There are in total three distinct homogeneous states, corresponding to $\rho_i^*=\{0,1,A\}$ for all $i$. A linear stability analysis, see hereafter for more details, shows that the homogeneous state $\rho_i^*=A$ is unstable while the two others are (locally) stable.

Any configuration in which stationary sites' densities are equal to $0$ or $1$ will be a fixed point of the system. There are $2^\Omega$ of such fixed points, including the two homogeneous states $\rho_i^*=\{0,1\}$  for all $i$. To determine their local stability, let us compute the Jacobian matrix $\mathbf{J}$ of the system. The $(i,j)$ element of this matrix is given by:
\begin{equation}
J_{ij} = \frac{A_{ij}}{4}\Big(g(\rho_i)-\rho_i g'(\rho_j)\Big) + \delta_{ij}\Big[f'(\rho_i)+ \sum_l \frac{A_{il}}{4} \Big(\rho_l g'(\rho_i)-g(\rho_l)\Big)\Big],
\end{equation}
with $A_{ij}=1$ if nodes $i$ and $j$ are nearest neighbors ($A_{ij}=0$ otherwise) \add{and the constant $D$ has been absorbed in the function $f$}.
Since $g(x)=x^a(1-x)$ (with $a>1$), it follows that $g(0)=0=g(1)$ and thus:
\begin{equation}
J_{ij}(\rho_k^*\in \{0,1\}) = -\frac{A_{ij}}{4} \rho_i^* g'(\rho_j^*) + \delta_{ij}\Big[f'(\rho_i^*)+ \sum_l \frac{A_{il}}{4}\, \rho_l^* g'(\rho_i^*)\Big],
\end{equation}
with $g'(x) = x^{a-1}\big[a-x(a+1)\Big]$. In particular, one has $g'(0)=0$ and $g'(1)=-1$. 

Let $i\in\{1,\cdots,\Omega\}$ be arbitrarily fixed. If $\rho_i^*=0$, then $J_{ij}=\delta_{ij}f'(0)$ for all $j=1,\cdots,\Omega$,  while if $\rho_i^*=1$, $J_{ij} = -\frac{A_{ij}}{4}g'(\rho_j^*) + \delta_{ij}\Big[f'(1)- \sum_l \frac{A_{il}}{4}\, \rho_l^*\Big]$, for all $j = 1,\cdots,\Omega$. Let us observe that $\sum_{j\neq i} \vert \frac{A_{ij}}{4}g'(\rho_j^*) \vert = \sum_{j\neq i} \frac{A_{ij}}{4}\rho_j^*$ \add{by using the fact that $g'(0)=0$ and $g'(1)=-1$}. By Gershgorin's theorem, we know that all the eigenvalues fall within the union of discs centered at $J_{ii}$ and of radius $R_i = \sum_{j\neq i}\vert J_{ij} \vert$. Since $f'(0)<0$ \add{and $f'(1)<1$},  we thus deduce\add{, by virtue of Gershgorin's theorem,} that all the eigenvalues lie in the complex half-plane and hence the fixed point is stable.

\section{Self-segregation in the slow-fast limit}\label{app:stabilityMatrix}
\begin{figure*}[htb!]
	\centering
	\begin{subfigure}
		\centering
		\includegraphics[width=0.45\textwidth]{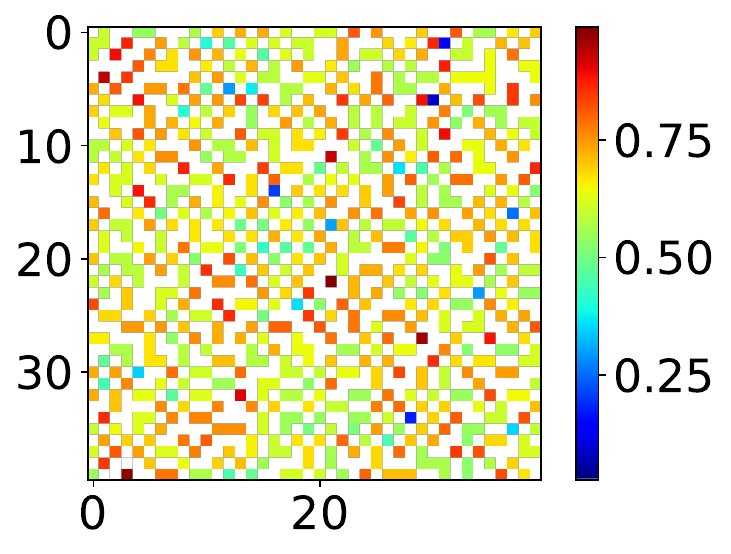}
		%\caption*{$\beta=0.3$}%
		%{{$\beta=0.34$}}    
		%\label{fig:mean and std of net14}
	\end{subfigure}
	\hfill
	\begin{subfigure}
		\centering
		\includegraphics[width=0.42\textwidth]{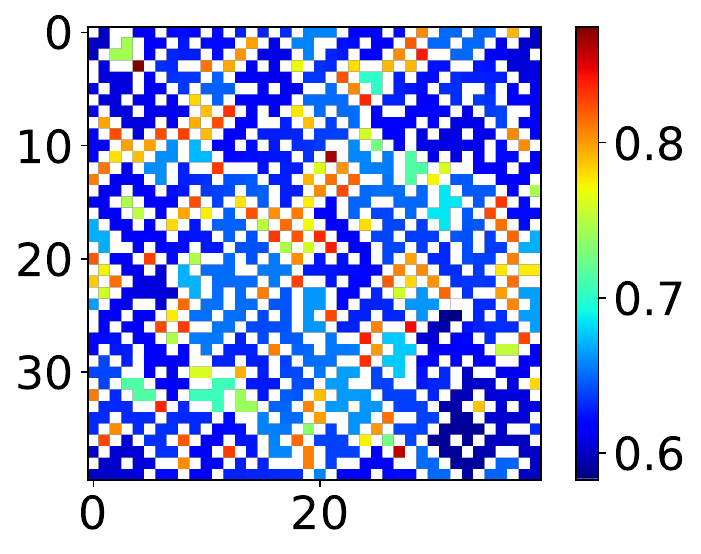}
		%\caption*{$\beta=0.4$}%
		%{{\small Network 1}}    
		%\label{fig:mean and std of net14}
	\end{subfigure}
	\caption{Stationary densities for a square lattice of dimension $40 \times 40$ and $g(x)=x^2(1-x)$. Initial densities were sampled in $[\beta-0.02,\beta+0.02]$, with $\beta_0=0.3$ (left) and $\beta_0=0.4$ (right).} 
	\label{fig:Diffusion}
\end{figure*}

\begin{figure*}[htb!]
	\centering
	\includegraphics[width=0.55\textwidth]{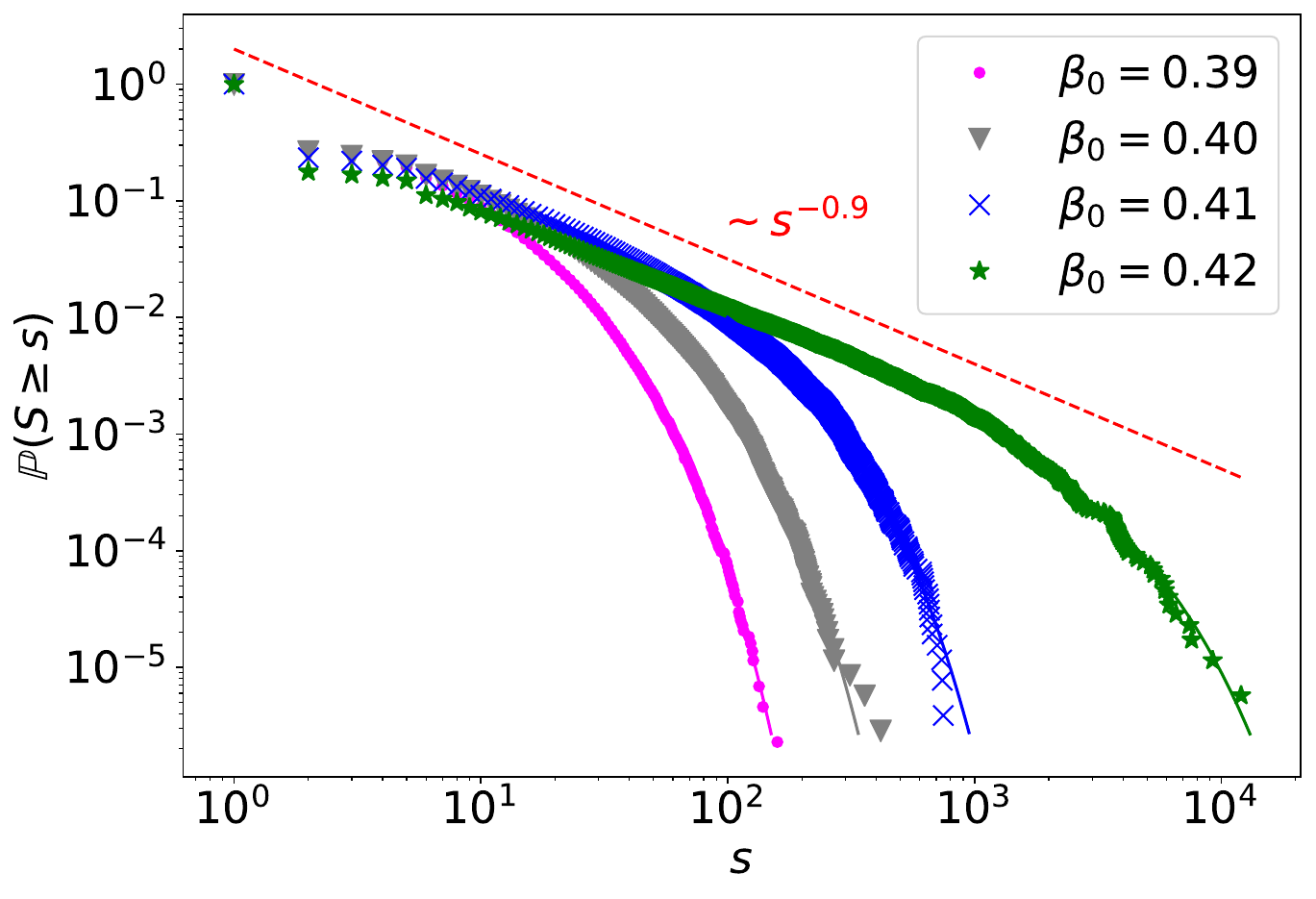}
	%\caption*{$\beta=0.3$}%
	%{{$\beta=0.34$}}    
	%\label{fig:mean and std of net14}
	\caption{The cumulative distribution \add{$\mathbb{P}(S\geq s)$} of cluster sizes is depicted, with each curve corresponding to a distinct average initial site density $\beta_0$ and representing the results of $10$ independent configurations. The data align well with a power-law distribution (indicated by the red dashed curve) that exhibits an exponential cutoff. The initial node densities were initialized within the range of $[\beta_0-0.01, \beta_0+0.01]$, followed by rescaling to ensure that the initial total mass divided by the number of sites equals $\beta_0$. The simulations were conducted on a square lattice of $500\times 500$ sites, with $a=2$.} 
	\label{fig:PLDiffusion}
\end{figure*}

In this section, we consider the limit $r\rightarrow 0$. In this case, the dynamical system boils down to the following equation:
\begin{equation}
\dot{\rho_i}= \sum_{j} \mathcal{L}_{ij} [\rho_j g(\rho_i)-\rho_ig(\rho_j)], ~i=1,\cdots,\Omega \, ,
\label{eqA:ODEs}
\end{equation}
with $\mathcal{L}_{ij} = \frac{A_{ij}}{k}-\delta_{ij}$. Following a linear stability analysis, see Sect.~\ref{app:fixedpoints}, we obtain that the homogeneous state $\rho_i^*=\beta$ is stable if and only if $\beta > \beta_c$, with
\begin{equation}
\beta_c = \frac{a-1}{a}\, .
\end{equation}  
Below this critical threshold, empty nodes emerge, as shown in Fig.~\ref{fig:Diffusion} where we report the stationary densities for $g(x)=x^2(1-x)$ and an average density $\beta=0.3$ (left panel) and $\beta=0.4$ (right panel). Further numerical analysis, see Fig.~\ref{fig:PLDiffusion}, indicates that the distribution of the cluster sizes is well-described by a power-law with an exponential cut-off, suggesting that \add{the distribution of mass into distinct clusters with power-laws scalings} is driven by diffusion.

\section{Intermittency in the self-segregation process}\label{app:Intermittency}

\begin{figure}
	\centering
	\includegraphics[width=.45\linewidth]{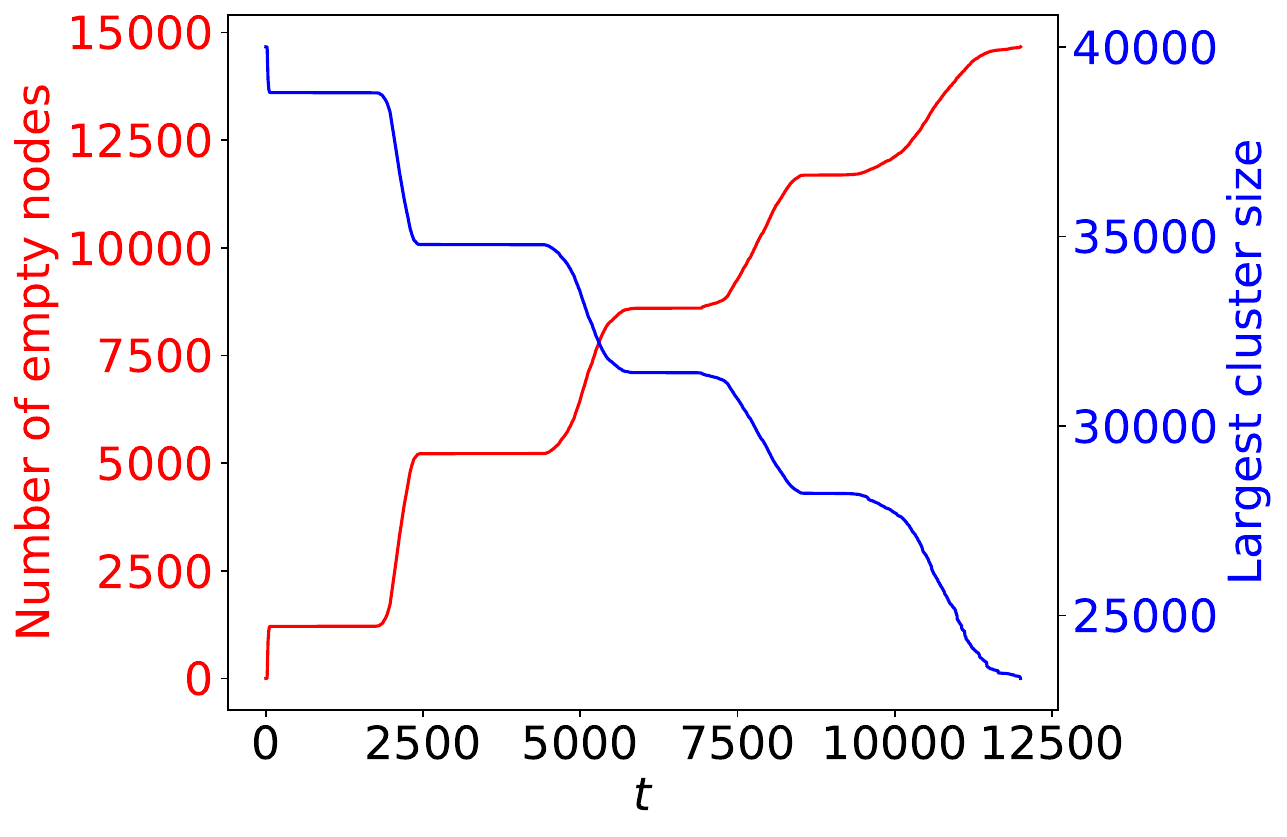}
	\hspace{1cm}
	\includegraphics[width=.45\linewidth]{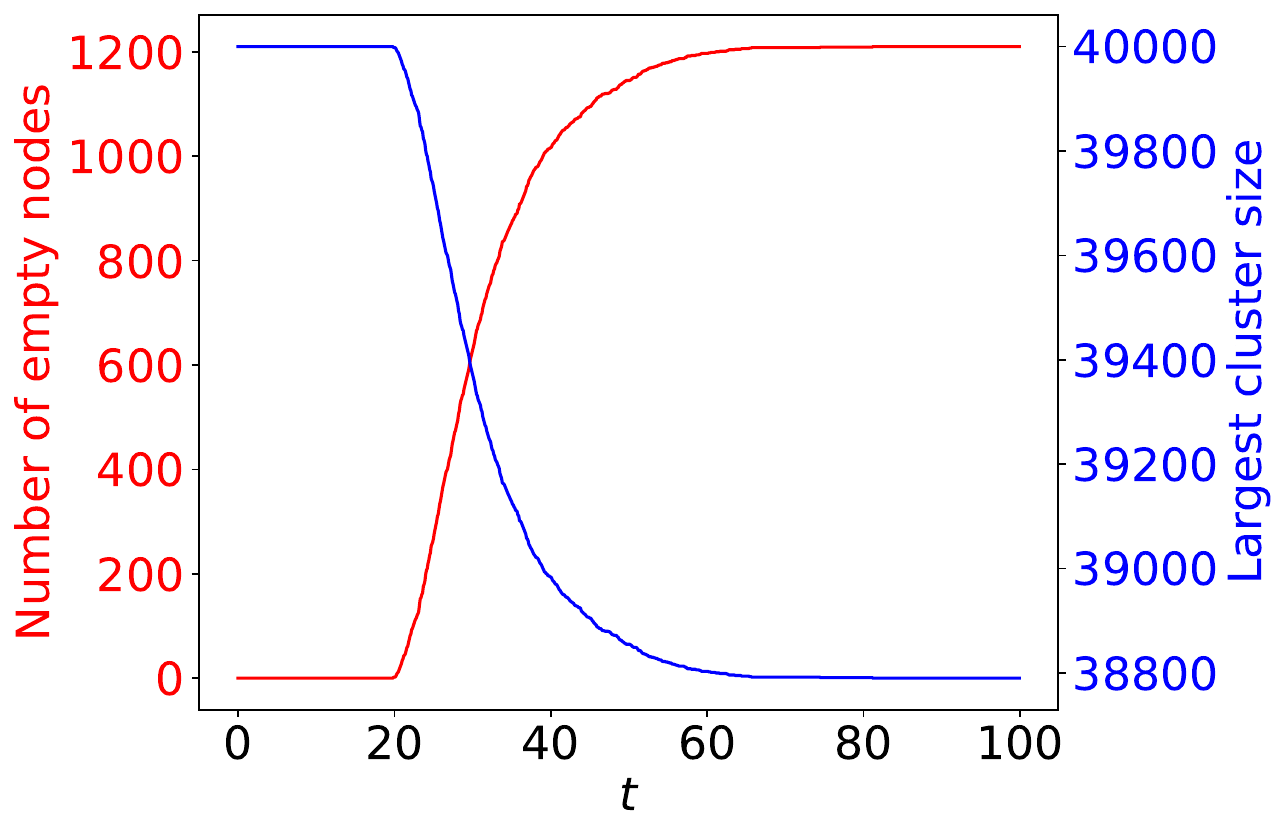}
	\caption{(Left) Number of empty nodes (red) and size of the largest cluster (blue) as a function of time for a square lattice of size $200 \times 200$. Initial conditions were randomly sampled in $[\beta_0-0.05,\beta_0+0.05]$ with $\beta_0 = 0.52$. Parameters are given by $r=0.001, D=10, A = 0.6$ and $g(\rho)=\rho^2(1-\rho)$. Nodes are considered empty if their density decreases below $\epsilon = 0.001$. (Right) Dynamics at early times, revealing a smaller plateau. }
	\label{fig:Intermittency}
\end{figure}

The dynamics in self-organized critical systems is characterized by large intermittent temporal phases \cite{pruessner2012self}. For instance, in the celebrated sandpile model, the slow addition of sand grains alternates with fast releases of sand, commonly called \textit{avalanches} \cite{pruessner2012self}. The self-segregation reaction-diffusion process considered in the main text, namely,
\begin{equation}
\frac{\partial \rho}{\partial t}=
r \rho (1-\rho) (\rho - A)
+ D \, \Big[
g(\rho)\Delta \rho - \rho \Delta g(\rho)
\Big]\ .
\label{eq:AppPDE}
\end{equation}
also exhibits intermittent phases in the slow fast setting, i.e., for $D/r \gg 1$. Fig. \ref{fig:Intermittency} (Left) shows the temporal evolution of the number of empty nodes as well as the size of the largest cluster, for $D/r = 10 000$ and a square lattice composed by $200 \times 200$ cells. The latter are considered empty if their density decreases below $\epsilon=0.001$. \add{Throughout this study,} a cluster refers to a set of nearest neighbor non-empty cells and the size of a cluster is the number of non-empty cells it contains. Let us observe that because of the dynamics Eq. (\ref{eq:AppPDE}), nodes will become empty or completely full only asymptotically, for this reason we have to fix a threshold \add{below which nodes are considered empty}. The size of the cluster will depend on such choice but not the general behavior resulting from intermittent dynamics. As it can be observed, the fraction of empty nodes and the size of the largest cluster correlate and manifest phases in which they remain constant (plateaus) interspersed with abrupt variations. The size of these plateaus also varies, as emphasized in Fig. \ref{fig:Intermittency} (Right) where the dynamics at early times reveals a smaller plateau. This provides further evidence that the self-segregation mechanism considered in this paper belongs to the class of self-organized processes \cite{pruessner2012self}.   

\end{document}